\documentclass[preprint,pra,aps,showpacs,preprintnumbers,amsmath,amssymb,eqsecnum]{revtex4}

\usepackage{graphicx}
\usepackage{dcolumn}
\usepackage{bm}
\usepackage{verbatim}
\usepackage{epsfig}
\usepackage{epstopdf}

\newcommand\beq{\begin{equation}}
\newcommand\eeq{\end{equation}}
\newcommand\bea{\begin{eqnarray}}
\newcommand\eea{\end{eqnarray}}

\def\s{{s}}
\def\a{\alpha}
\def\b{\beta}
\def\av{{\bf a}}
\def\bv{{\bf b}}

\def\half{\frac {1} {2}}

\def\x0{{{\bf x}_0}}

\def\an{{\a_1, \a_2 \cdots \a_n}}
\def\bn{{\b_1,\b_2 \cdots \b_n}}

\begin{document}



\title{Negative Probabilities, Fine's Theorem and Linear Positivity}

\author{J.J.Halliwell}%
\email{j.halliwell@imperial.ac.uk}

\affiliation{Blackett Laboratory \\ Imperial College \\ London SW7
2BZ \\ UK }

\author{J.M.Yearsley}
\email{jmy27@cam.ac.uk}
\affiliation{Centre for Quantum Information and Foundations, DAMTP, Centre for Mathematical Sciences, University of Cambridge, Wilberforce Road, Cambridge CB3 0WA, UK}



\begin{abstract}
Many situations in quantum theory and other areas of physics lead to quasi-probabilities
which seem to be physically useful but can be negative. The interpretation of such
objects is not at all clear. In this paper, we show that quasi-probabilities naturally
fall into two qualitatively different types, according to whether their non-negative
marginals can or cannot be matched to a non-negative probability. The former type,
which we call viable, are qualitatively similar to true probabilities, but the latter
type, which we call non-viable, may not have a sensible interpretation.
Determining the existence of a probability matching given marginals is a non-trivial question in general. In simple examples, Fine's theorem indicates that inequalities of the Bell and CHSH type provide criteria for its existence, and these examples are considered in detail.
Our results have consequences for the linear positivity condition of Goldstein and Page in the context of the histories approach
to quantum theory. Although it is a very weak condition for the assignment of probabilities it fails in some important cases where our results indicate that probabilities clearly exist. We speculate that our method, of matching probabilities to a given set of marginals, provides a general method of assigning probabilities to histories and we show that it passes the Di\'osi test for the statistical independence of subsystems.

\end{abstract}

\pacs{03.65.Yz, 03.65.Ta, 02.50.Cw}


\maketitle

\section{Introduction}

Many situations in physics and mathematics invite the extension of the notion of probability to quasi-probabilities whose values may be negative, or greater than $1$. Quantum theory in particular frequently
leads to such distributions when non-commuting operators are involved.
The Wigner function
is a particularly obvious example \cite{Wig}.
It is often claimed that despite being sometimes negative, quasi-probabilities can contain a lot of intuitively useful information about the physical system, a position argued perhaps most eloquently by Feynman \cite{Fey}.
However, in the absence of Feynman's intuition, the realm of quasi-probabilities
runs some risk of becoming a lawless land
%
and there are surely some limits
to their sometimes indiscriminate interpretation.

In this paper, we make a number of observations about quasi-probabilities and, in particular, we offer a concrete proposal for assessing and possibly restricting their use. Our remarks will apply to any quasi-probability, but we will focus on quasi-probabilities of the type that arise in quantum mechanics when non-commuting operators are involved. There are two important types of examples of such quasi-probabilities. The first is the Wigner function, in both its continuous \cite{Wig} and discrete form \cite{Woo}. It is essentially the discrete version that Feynman discussed \cite{Fey}, and this was followed up in Ref.\cite{SWS}. The second example is the formula used in the linear positivity condition of Goldstein and Page \cite{GoPa,Dio,Hal0}, which is the real part of an average of non-commuting projectors, discussed further by Hartle \cite{Har2,Har}. (See also the more elaborate approach to quantum theory with extended probabilities by Gell-Mann and Hartle \cite{GeHaReal}).

These examples indicate that when non-commuting operators are involved, there are no {\it unique} prescriptions as to which quasi-probability formula is the most appropriate
for a given situation. Different operators orderings will yield similar formulae which may be just as useful. This means that it is reasonable to explore families of quasi-probabilities
with similar properties.
In particular, we explore the possibility that a given quasi-probability can be related in some way to a non-negative probability which shares with the quasi-probability as many characteristics as possible.
This could be done, for example, by considering modifications of the original quasi-probability.



The key question is which characteristics of the probability and quasi-probability should be the same? Our specific proposal is to consider the marginals of the quasi-probability. Although the quasi-probability is sometimes negative it will always have some positive marginals. The examples mentioned above have precisely this property.
We thus look for a positive probability whose marginals match the positive marginals
of the quasi-probability. The resulting positive probability is then interpreted
in the usual way.

However, finding a probability matching a given set of marginals
is a non-trivial problem in general and, crucially, is often not possible \cite{Fine1,Fine2,GaMer,Pit}. This means that there is a potentially significant way of dividing quasi-probabilities into two groups: those that can be related to a positive probability with some of the same characteristics, and those that cannot, since their positive marginals cannot be matched to a positive probability. Will shall refer to the former group as ``viable'' and the latter, ``non-viable".

This division into viable and non-viable is perhaps important because, in our view,
quasi-probabilities are often used
{\it as if} there was an associated probability and this is perhaps the tacit assumption behind their intuitive interpretation. It is therefore important to determine if
this tacit assumption is really true in each case.
When the quasi-probability is non-viable, an associated probability with similar characteristics does not exist, which throws into question its interpretation. Viable quasi-probabilities, on the other hand, may avoid issues with their interpretation since they are associated with a true probability. Moreover, the viability test can also be useful in identifying the existence of a relevant probability even when standard formulae for quasi-probabilities, such as linear positivity or the Wigner function, are negative, so do not signal the possible existence of a probability.

To be clear, we are not in this paper offering a new interpretation of quasi-probabilities.
Rather, we are suggesting that there are some systematic notions and reasonable
restrictions behind their intuitive interpretation.


Many previous authors have addressed the question of making sense of quasi-probabilities. See for example Refs.\cite{Har2,Har,SWS,Cin,Kro,You,Bur} and also the review Ref.\cite{Muc}.
Of particular relevance to the present work are those previous works which explore quasi-probabilities in the context of EPRB correlations \cite{Wod,Cer,HHK,RoSu,HLS}. The present work also has some bearing on the decoherent histories approach to quantum theory \cite{GH1,GH2,GH3,Gri,Omn1,Omn2,Hal2,Hal3,DoK,Ish,IshLin} and in particular, as mentioned, the linear positivity condition of Goldstein and Page \cite{GoPa,Dio,Hal0,Har2,Har}.

In Section 2, we set out our proposal in detail. In Section 3 we consider
some simple models in which our proposal can be carried out
in detail. We generalize these models in Section 4. We consider the EPRB situation in Section 5 and we
compare with the linear positivity condition of Goldstein and Page. In Section 6 we introduce some more speculative ideas about the use of our methods in a more general way for the assignment of probabilities to histories.
We summarize and conclude in Section 7.

In what follows, the difference between
``positive" and ``non-negative" probabilities is not important, so for ease of language
we will simply use the word ``positive" in an looser way so that it includes ``non-negative''.




\section{The Proposal}

Our proposal for assessing the viability of quasi-probabilities is not specifically tied to quantum theory but to fix ideas it is most useful to set up the problem in the
quantum context. We therefore suppose we are interested
in a set of probabilities or quasi-probabilities which refer to a set of hermitian
operators, $A_1 \cdots A_n$, some of which do not commute with each other. These
could, for example, be the spins in various directions in a system of one or two
particles, or the positions at different times for a point particle. For
each operator $A_k$, one can construct a projection operator $P_{\a_k}^k$ onto a
range of its spectrum, labeled by alternative $\a_k$, which satisfy
\beq
\sum_{\a_k} P_{\a_k}^k = 1
\eeq
If the initial state is $\rho$,
the probability for the alternative $\a_k$ is given by
\beq
p(\a_k) = {\rm Tr} \left( P_{\a_k}^k \rho \right)
\label{2.1}
\eeq
where $\rho$ is the initial density operator.
This probability is clearly positive.

Since the set of operators do not commute, there
is no obvious unique probability formula for the full set of alternatives, $\a_1 \cdots \a_n$.
The decoherent histories approach gives one possible approach to this issue
\cite{GH1,GH2,GH3,Gri,Omn1,Omn2,Hal2,Hal3,DoK,Ish,IshLin} but works with candidate probabilities which are always positive. Here, we are interested instead in quasi-probabilities and a useful example to focus on is
\beq
q(\a_1, \cdots \a_n) = {\rm Re} {\rm Tr} \left( P_{\a_n}^n \cdots P_{\a_2}^2 P_{\a_1}^1
\rho
\right)
\label{qformula}
\eeq
It clearly satisfies
\beq
\sum_{\a_1 \cdots \a_n} q (\a_1, \a_2, \cdots \a_n) = 1
\eeq
since the projectors sum to $1$.
This quasi-probability is not in general positive but will be if all the projectors commute.  Furthermore, it can be positive for some initial states and some projection
operators, but this has to be checked in particular models. This is the basis of the linear positivity condition of Goldstein and Page \cite{GoPa,Dio,Hal0} and will be discussed below.
Quasi-probabilities with similar properties are obtained using
different operator orderings of the projectors in Eq.(\ref{qformula}). Quasi-probabilities of a somewhat different type, but with marginals of the form Eq.(\ref{2.1}), may also be obtained using constructions of the Wigner function type \cite{Wig,Woo}.

If $n-1$ alternatives are summed out of $q (\an)$, the single-alternative marginals of the form Eq.(\ref{2.1}) are obtained and these are positive. Generally, all marginals of Eq.(\ref{qformula}) corresponding to commuting sets of operators are necessarily positive for all initial states. Furthermore, some of the marginals involving non-commuting operators may
be positive, depending on the initial state and specific choices of projections,
in some models.

The situation outlined above depicts the general situation we are interested in.
We suppose we have a set of quasi-probabilities $ q (\a_1, \a_2, \cdots \a_n) $ which sum to $1$. They may lie outside the range $[0,1]$ but some of their marginals will
take values in the range $[0,1]$.
We are interested in associating the quasi-probabilities $q (\an) $ with a true probability
$p (\a_1, \a_2, \cdots \a_n)$, so it also sums to $1$ but always takes values in $[0,1]$.
The key idea is to consider the marginals of $q (\an) $. Our proposal is then
the following: {\it Find all the marginals of $q (\an) $ which lie in the range $[0,1]$ and look
for a positive probability $p (\an )$ which matches those marginals}.
As stated earlier, if a probability can be found, we will refer
to the original quasi-probability as viable, otherwise it is non-viable.
We will refer to the above procedure as the ``viability test'' for quasi-probabilities.
(In what follows, for simplicity of language, we will focus on the possible negativity of quasi-probabilities and decline from explicitly referring to quasi-probabilities which 
are greater than $1$ -- theses two properties are clearly connected by normalization).

The question now is to what extent this programme can be carried out in particular
examples. Finding a probability which matches the one-alternative marginals (which we assume
are positive) is easy. For example,
\beq
p(\an ) = q(\a_1) q(\a_2) \cdots q (\a_n)
\label{onetime}
\eeq
does the job, although this is not the only solution.
The first non-trivial situation arises when some of the two-alternative marginals,
of the form $q(\a_j, \a_k)$ for some $j,k$, are positive. We then seek a positive
probability $p(\an)$ matching these marginals. This is a non-trivial problem in general
although in simple cases it is related to the CHSH inequalities \cite{Fine1,Fine2,GaMer,Pit,CHSH}, as we shall discuss in the next section. We anticipate that more complicated cases
will be difficult to solve, unless they have simplifying features. However, it is
sometimes possible to prove that a quasi-probability is non-viable in some quite
general cases, as we will outline below.

We now offer some further remarks as to the significance of the division of quasi-probabilities
into viable and non-viable. This division, recall, was partly motivated by the observation
that there is no unique quasi-probability associated with a set of non-commuting
operators so it is reasonable to explore families of quasi-probabilities. A viable
quasi-probability is one in which there are a family of quasi-probabilities with
the same positive marginals, and some of these quasi-probabilities
are positive, so it
is reasonable to switch attention to the positive ones.
Loosely, it means a positive probability exists but we have picked the ``wrong'' formula.
We could also say that for a viable quasi-probability $q(\an)$, there exists a set
of numbers $ d(\an )$ such that
\beq
p(\an ) = q(\an ) + d (\an)
\label{mod}
\eeq
is positive and has the same positive marginals as $q(\an)$. That is, a positive probability
is obtained by a simple modification of $q(\an)$ which preserves the positive marginals.
The modification $d (\an)$ is in general not unique which means that it may be desirable
to impose further conditions. One natural condition is to require that if any
one of the $ q(\an)$'s is zero, then the corresponding $p(\an)$ should be set to
zero. Note also that there is in this modification no requirement of ``closeness'' of $ p(\an)$ and $q(\an)$. This could be considered, but we will not do this here. Since viable quasi-probabilities are linked to a true probability, it seems reasonable
to conjecture that their intuitive interpretation will not lead to physically unreasonable
results.  In simple quantum-mechanical models of the type considered in the following sections, the notion of viable quasi-probability is related to the idea of an underlying hidden variables theory.

A non-viable quasi-probability, by contrast, is one for which there is no associated
probability and there exists no modification $d(\an)$ which would produce a positive
probability whilst preserving the positive marginals. This means that it is
not consistent to assert that there exist a set of underlying quantities which possess
values. We speculate, but have not proved, that manipulating the marginals of a
non-viable quasi-probability as if there was an underlying probability may lead to physically unreasonable results. This is
the case in some simple quantum-mechanical models, where non-viability is related to situations in which there is no underlying hidden variables theory.

In simple terms, viable quasi-probabilities are qualitatively similar to probabilities,
but non-viable quasi-probabilities are not. These remarks will be illustrated in the examples of the following sections.

\section{Some Simple Models}

A particularly instructive example of our ideas, already the focus of much study,
is the situation in which there are four alternatives, $\s_1, \s_2, \s_3, \s_4$
taking values $ \pm 1$. We suppose that we are given a quasi-probability
$ q (\s_1, \s_2, \s_3, \s_4) $ for which the four marginals
$ q( \s_1, \s_3)$, $ q (\s_1, \s_4)$, $q (\s_2, \s_3)$ and $ q (\s_2, \s_4) $ are positive.
We wish to modify it into a positive probability $ p (\s_1, \s_2, \s_3, \s_4 )$ using the strategy outlined above.
We thus seek a positive probability $ p (\s_1, \s_2, \s_3, \s_4)$ which matches the given marginals.  This is of course the situation encountered in the CHSH analysis of the EPRB pair \cite{Fine1,Fine2,GaMer,Pit,CHSH} (and similarly in the Leggett-Garg analysis of temporal correlation functions
\cite{LeGa}) but for now we keep the analysis general and not tied to a specific application.

There is a particularly useful form of the quasi-probability, which is its expansion
in terms of its correlation functions,
\beq
q (\s_1, \s_2, \s_3, \s_4) = \frac {1} {16} \left( 1 +
\sum_i B_i s_i + \sum_{ij} C_{ij} s_i s_j + \sum_{ijk} D_{ijk} s_i s_j s_k
+ E s_1 s_2 s_3 s_4 \right)
\label{qs}
\eeq
where the indices $i,j,k$ run over the values $1,2,3,4$, the summation over $ij$ has $i \ne j$ and the summation over $ijk$ has all three indices different.
The coefficients are given by
\bea
B_i &=& \sum_{\s_1 \s_2 \s_3 \s_4} \ \s_i \ q (\s_1, \s_2, \s_3, \s_4 )
\nonumber \\
C_{ij} &=& \sum_{\s_1 \s_2 \s_3 \s_4} \ \s_i \s_j \ q (\s_1, \s_2, \s_3, \s_4)
\nonumber \\
D_{ijk} &=& \sum_{\s_1 \s_2 \s_3 \s_4} \ \s_i \s_j \s_k\  q (\s_1, \s_2, \s_3, \s_4)
\nonumber \\
E &=& \sum_{\s_1 \s_2 \s_3 \s_4}  \ \s_1 \s_2 \s_3 \s_4 \ q (\s_1, \s_2, \s_3, \s_4 )
\eea
The marginals are then easily constructed by summing out some of the $\s_i$'s. So
for example
\beq
q(\s_1, \s_3) = \frac{1} {4} \left( 1 + B_1 \s_1 + B_3 \s_3 + C_{13} \s_1 \s_3 \right)
\eeq
This means that fixing the given four marginals is equivalent to fixing the values of $B_i$ for $i=1,2,3,4$
and the values of the correlation functions $C_{13}$, $C_{23}$, $C_{14}$ and $C_{24}$.
(That is, Eq.(\ref{qs}) shows that it is always possible to find a quasi-distribution
which matches the given four marginals, a fact which has been previously noted \cite{Wod,Cer,RoSu,HLS}.)
The question of finding a positive probability is then the question of whether the
remaining unfixed correlation functions, $ C_{12}$, $C_{34} $, $D_{ijk}$ and $E$ can be chosen in such a way that the quasi-probability Eq.(\ref{qs}) is positive.

The answer to this question is contained in an important theorem due to Fine \cite{Fine1,Fine2}.
First of all, it is well-known, and easily shown, that if a probability
exists, then the correlation functions must obey the eight CHSH inequalities \cite{CHSH}
\bea
-2 & \le & C_{13} + C_{14} + C_{23} - C_{24} \le 2
\\
-2 & \le & C_{13} + C_{14} - C_{23} + C_{24} \le 2
\\
-2 & \le & C_{13} - C_{14} + C_{23} + C_{24} \le 2
\\
-2 & \le & - C_{13} + C_{14} + C_{23} + C_{24} \le 2
\eea
Fine's theorem says that the converse is also true:
{\it If} the eight CHSH inequalities hold, then there exists a probability distribution
$p (\s_1, \s_2, \s_3, \s_4)$ matching the given marginals. That is, if the CHSH inequalities
hold, then values of $ C_{12}$, $C_{34} $, $D_{ijk}$ and $E$ may be chosen in such
a way that the probability is positive.

The proof of Fine's theorem is not trivial. Fine gave a direct proof in Refs.\cite{Fine1,Fine2}
by showing, at some length, by purely algebraic means how to flesh out the given set of marginals
into a full probability distribution. A more insightful proof, using the
geometry of polytopes, was given by Pitowski \cite{Pit}. Another proof is contained in the
work of Garg and Mermin, who considered a general class of problems of
this type \cite{GaMer}.


Fine's theorem and the CHSH inequalities supply the conditions under which the quasi-probability
in this example may be linked to a true probability. If the CHSH inequalities are not satisfied, there is no associated probability matching the given marginals.
Hence, the CHSH inequalities give a concrete set of criteria for ruling out the viability of certain quasi-probabilities.

If the CHSH inequalities are satisfied, then there is in general
a family of probabilities matching the given
marginals, so some extra conditions may be required to reduce their number.
This family is parametrized by the freedom remaining in the
correlation functions $ C_{12}$, $C_{34} $, $D_{ijk}$ and $E$ once the CHSH inequalities
are satisfied. It could be further restricted in various ways as discussed in the previous
section.

Note also that the CHSH inequalities do not force {\it every} quasi-probability matching
the given marginals to be positive. When the CHSH inequalities are satisfied there will be both probabilities {\it and} quasi-probabilities matching the given marginals.
This illustrates the general statement given in the previous section: the existence
of quasi-probabilities is sometimes indicative of a family of quasi-probabilities
with similar characteristics, some of which are true probabilities, so it is natural
to switch attention to the probabilities.

Some related, but different, observations about quasi-probabilities for this system were made in Refs.\cite{Wod,Cer,RoSu,HLS}. Not all of these authors were aware of Fine's theorem and its relevance to the situation.


We also note that there is a slightly simpler example that could also be useful,
namely the case in which there is a quasi-probability $ q (\s_1, \s_2, \s_3) $, with $\s_i = \pm 1$, for
which the three marginals $ q (\s_1, \s_2)$, $q(\s_2, \s_3)$ and
$q (\s_1, \s_3)$ are positive. The quasi-probability in this case may be written
\beq
q (\s_1, \s_2, \s_3) = \frac {1} {8} \left( 1 +
\sum_i B_i s_i + \sum_{i \ne j} C_{ij} s_i s_j + D s_1 s_2 s_3 \right)
\label{qs2}
\eeq
where $i,j,k$ runs over values $1,2,3$. In this case the necessary and sufficient conditions
for the existence of a probability are the Bell inequalities \cite{Pit}
\bea
C_{12} + C_{13} - C_{23} \le 1
\\
C_{12} - C_{13} + C_{23} \le 1
\\
- C_{12} + C_{13} + C_{23} \le 1
\\
- C_{12} - C_{13} - C_{23} \le 1
\eea

Refs.\cite{Fey,SWS} involved the even simpler example of a quasi-probability $q(\s_1, \s_2)$, with $\s_i = \pm 1$, matching the marginals $q(\s_1)$, $q(\s_2)$. This is easily shown to be viable.

\section{More General Cases}

Although the above examples are very useful since they can be carried through in detail,
they do not obviously extend to any more general cases. The question of finding probability
distributions matching given marginals is a very difficult problem in general \cite{GaMer}.
However, the simple examples above can be used to prove that more general quasi-probabilities are non-viable, using a coarse graining procedure.
The point is that if a quasi-probability is viable then all coarse grainings of it must also be viable.

In the general case
we seek a probability $p (\an)$ matching
the positive marginals of a quasi-probability $q(\an)$, where each $\a_j$ runs over $d_j$ values.
By summing over ranges of each $\a_j$ or by summing them out altogether in $q(\an)$,
we can easily obtain quasi-probabilities of the form $ q( \s_1, \s_2, \s_3, \s_4)$,
where $\s_i = \pm 1$.
We can in a similar way coarse grain the specified positive marginals of $ q (\an)$.
Depending on exactly which marginals are specified, we will at least in some cases, obtain
the marginals specified in the CHSH or Bell case.

This coarse graining down to a CHSH or Bell problem can of course be done in a number of different ways. In each case, we can then ask if there is an associated probability, using the CHSH or Bell inequalities.
The point now is that if the CHSH or Bell inequalities fail for any one of them, then the
original quasi-probability $q( \an)$ is non-viable. If however, the CHSH inequalities are satisfied for all such coarse grainings, then this
procedure does not tell us anything about $ q (\an)$ -- it may or may not be viable.

As stated, our definition of viable probabilities has obvious connections with the existence of hidden variable theories in quantum-mechanical models. A particularly relevant hidden variable theory is Bohmian mechanics \cite{Bohm,BoHi,Holl}, not least since it easily invites comparison with the decoherent histories approach and linear positivity \cite{HaBohm,GriBohm}. In Bohmian mechanics the probabilities for histories are always non-negative and match the single time probabilities supplied by decoherent histories or linear positivity. The probabilities for two or more times do not match in general and can be very different \cite{HaBohm,GriBohm}, but they can also be very similar under quasi-classical conditions. These remarks suggest that it would be of particular interest to explore the connection between viable set of probabilities and those supplied by Bohmian mechanics. This will be explored in another paper.

\section{EPRB and Linear Positivity}

In the context of the consistent histories approach to quantum theory \cite{GoPa},
Goldstein and Page have proposed to use the formula Eq.(\ref{qformula}) to assign
probabilities to histories. It has the advantage that it satisfies all the probability
sum rules, but it is not always positive. They therefore proposed that one focuses on situations in which  Eq.(\ref{qformula}) is positive. This condition, which is known as linear positivity, is the weakest condition proposed so far for the assignment of probabilities to histories (or more generally, to non-commuting variables), in the context of the decoherent histories approach to quantum theory
\cite{GH1,GH2,GH3,Gri,Omn1,Omn2,Hal2,Hal3,DoK,Ish,IshLin}.

In the Goldstein-Page scheme, situations for which  Eq.(\ref{qformula}) is negative
are rejected. However,
in the framework presented in this paper, such quasi-probabilities are not automatically
rejected, but need to be carefully examined to see if they are viable. We anticipate
that there will, in fact, be numerous situations in which linear positivity fails,
but the quasi-probabilities are still viable. This means that the Goldstein-Page scheme could then lead to a positive probability if modifications of the quasi-probability
of the form Eq.(\ref{mod}) were considered.

To show this, we consider a particular example in detail, namely the CHSH analysis
of the EPRB state \cite{CHSH}. We will exhibit a situation in which the linear positivity condition
fails, but the CHSH inequalities are satisfied, so that an underlying probability exists.

The situation is the standard EPRB set up, in which
we consider a pair of particles $A$ and $B$ whose spins are in the singlet state,
\beq
| \Psi \rangle = \frac {1} {\sqrt{2}} \left( | \uparrow \rangle \otimes | \downarrow \rangle
- | \downarrow \rangle \otimes | \uparrow \rangle \right)
\eeq
where $ |\uparrow \rangle $ denotes spin up in the $z$-direction.  Measurements are
made on particle $A$ in the directions characterized by unit vectors ${\bf a}$ and
${\bf a}'$ and on particle $B$ in directions ${\bf b}$ and $ {\bf b}'$.
The probabilities for pairs of such measurements, one on $A$, one on $B$ is of the
form
\beq
p(s_a, s_{b}) = {\rm Tr} \left( P_{s_a}^{\bf a} \otimes P_{s_b}^{\bf b} | \Psi \rangle
\langle \Psi | \right)
\eeq
Here, $s = \pm 1 $ and the measurements are described by projection operators of the form
\beq
P_{s}^{\bf a} = \half \left( 1 + s {\bf a} \cdot \sigma \right)
\eeq
where $\sigma_i$ denotes the Pauli spin matrices. With the two different directions for the measurements on two particles, there are four pairs of such measurements with positive probabilities.

\begin{figure}[h]
\begin{center}
\includegraphics[width=5in]{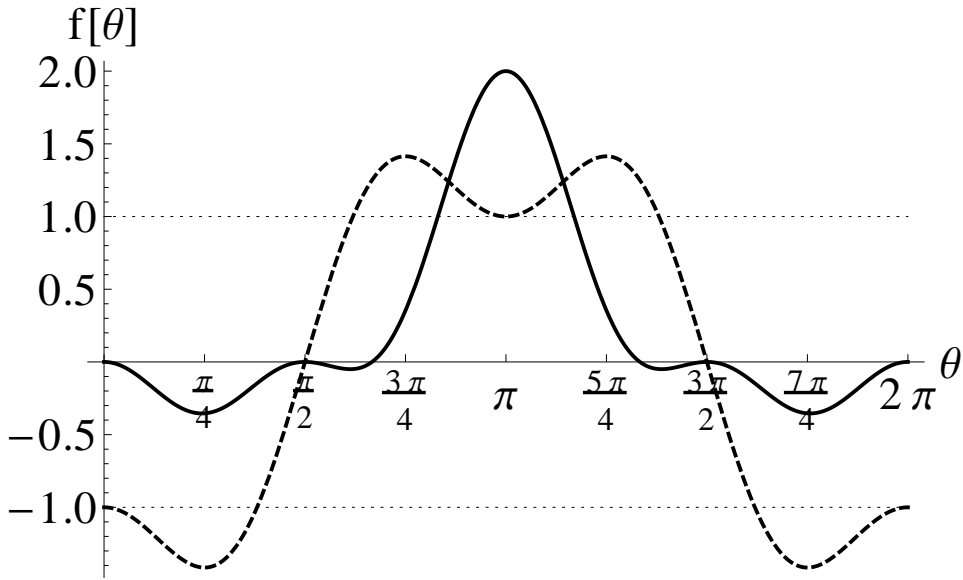}
\caption{The bold line shows the probability $q(++++)$ (which is scaled by a factor of $4$ for clarity) and the dashed line shows CHSH the inequality Eq.(\ref{CHSH1}). There are clear regions around $\theta = \pi /2 $ and
$ \theta = 3 \pi / 2 $ where the CHSH inequalities are satisfied but the quasi-probability is negative.}
\label{fig1}
\end{center}
\end{figure}

We are interested in the quasi-probability
\beq
q (\s_1, \s_2, \s_3, \s_3 ) =  {\rm Re} {\rm Tr} \left( P_{\s_1}^{\bf a} P_{\s_2}^{\bf a'} \otimes P_{\s_3}^{\bf b} P_{\s_4}^{\bf b'} | \Psi \rangle \langle \Psi | \right)
\label{5.4}
\eeq
This is not positive in general but
matches the positive marginals of interest
$ q( \s_1, \s_3)$, $ q (\s_1, \s_4)$, $q (\s_2, \s_3)$ and $ q (\s_2, \s_4) $. The Goldstein-Page scheme is to look for situations
in which Eq.(\ref{5.4}) is positive. By explicit calculation it is readily shown that
\bea
q (\s_1, \s_2, \s_3, \s_3 ) &=& \frac{1} {16} \left[ ( 1 + \s_1 \s_2 \av \cdot \av')
( 1 + \s_3 \s_4 \bv \cdot \bv') - (\s_1 \av + \s_2 \av')\cdot (\s_3 \bv + \s_4 \bv')
\right.
\nonumber \\
&+& \left. \s_1 \s_2 \s_3 \s_4 \left( \av \cdot \bv \ \av' \cdot \bv' - \av \cdot \bv' \ \av' \cdot \bv\right) \right]
\eea
(which is clearly of the form Eq.(\ref{qs})).
 We choose the orientation of the four vectors so that they are coplanar with
\bea
\av \cdot \bv &=& \av \cdot \bv' = \av' \cdot \bv = \cos \theta
\nonumber \\
\av' \cdot \bv' &=& \cos 3 \theta
\label{5.theta}
\eea
where $0 \le \theta \le 2 \pi $.
Consider the component of the quasi-probability
in which $ \s_i = +1 $ for all $i$. It is straightforward to show that
\beq
q(++++) =  \frac{1} {4} \cos^2 \theta \left( 2 \cos^2 \theta  - \cos \theta - 1 \right)
\eeq
This is plotted in Fig.(\ref{fig1}).
It is negative for a range of values of $\theta$ so linear positivity fails.

Now we check for the viability of this quasi-probability.
The CHSH inequalities in this case take the familiar form,
\beq
-2 \le \av \cdot \bv + \av' \cdot \bv + \av \cdot \bv' - \av' \cdot \bv' \le 2
\eeq
plus three more similar expression with the minus sign in the other three possible locations. With the choice of vectors given above,
the CHSH inequalities then reduce to just two conditions, namely,
\bea
&-1& \le 2 \cos^3 \theta - 3 \cos \theta \le 1
\label{CHSH1}
\\
&-1& \le 2 \cos^3 \theta -  \cos \theta \le 1
\eea
The second inequality is always satisfied.
The first is satisfied for a wide range of values, in particular for
$ \theta $ in a neighbour around $ \pi/2$ and around $ 3 \pi / 2$, regions where $q(++++)< 0$. This is depicted in Fig.(\ref{fig1}).
By Fine's theorem, discussed above, this therefore means that a probability
$ p(\s_1, \s_2, \s_3, \s_4) $ exists which matches the given marginals
so the quasi-probability is viable.

In brief, there is an underlying probability in this situation, but it is not the
one supplied by the Goldstein-Page scheme, since the linear positivity condition
fails. This is a key example which shows that it is clearly physically reasonable
to search for a true probability to describe the situation even when the obvious quasi-probability is sometimes negative.

Some further comments are appropriate. First of all, linear positivity must fail
if the CHSH inequalities are not satisfied, since in that case there cannot be any
probability matching the given marginals. Secondly, one could consider modifications
of the quasi-probability Eq.(\ref{5.4}) involving different operator orderings of
the projectors.
We have checked some of these and found similar conclusions.
(Some similar calculations
were carried out in Ref.\cite{RoSu} but these were focussed on the different issue of linking negative quasi-probabilities to violations of the CHSH inequalities.)

An interesting question is whether
there is a general quasi-probability of the form Eq.(\ref{5.4})
which is positive if and only if the CHSH inequalities hold. We have not been able to answer this question. It is not obvious from the detailed proofs of Fine's theorem \cite{Fine1,Fine2,Pit,GaMer}. We suspect there is no such general formula -- the probability exists when the CHSH inequalities hold but its specific form probably depends in some detail on the values of the given correlation functions. This will be explored elsewhere.


We finally note a slightly curious property of the above example. Linear positivity requires that all sixteen components of the quasi-probability Eq.(\ref{5.4}) are non-negative. However, with the
choice of orientation Eq.(\ref{5.theta}) it is not hard to show that this is true only at isolated points, for example, $\theta = \pi/2 $. This means that linear positivity is not satisfied in a robust way, with this choice, although viability clearly is. We imagine that linear positivity could be satisfied more robustly with other choices of orientation of the vectors but we have not explored this. If anything, this feature emphasizes our conclusion: viability is a weaker condition than linear positivity.

\section{A General Method of Assigning Probabilities to Histories and the Di\'osi Test}

We now make some speculative remarks on the possible application of our ideas to histories-based approaches to quantum theory.  Our procedure, of finding probabilities matching a given set of marginals, provides a way of assigning families of probabilities to histories that might be useful even if the linear positivity condition is satisfied.

Suppose one is interested in finding a set of probabilities for histories matching a given set of marginals. The marginals could, for example, be those corresponding to the commuting subsets of variables, for which all approaches agree how to assign probabilities. Linear positivity, when satisfied, gives one way, via a specific formula, to assign probabilities. Our procedure gives a way of finding a whole family of probabilities, which will include the Goldstein-Page formula, Eq.(\ref{qformula}), as a member of the family. However, other members of the family might be regarded as equally valid candidate probabilities, given the non-uniqueness associated with assigning probabilities to non-commuting variables.

This broadening of the definition of probabilities for histories may be useful for the following reason. It was pointed out by Di\'osi that linear positivity (and also some of the other weaker decoherence conditions) suffer from a serious problem when applied to uncorrelated composite systems \cite{Dio} (see Ref.\cite{Hal0} for further discussion). In brief, the linear positivity condition may be written
\beq
{\rm Re} {\rm Tr} \left( C_{\a} \rho \right) \ge 0
\label{LP1}
\eeq
where the class operator $C_{\a}$ denotes the string of projectors in Eq.(\ref{qformula}). However, for a composite system of two uncorrelated components with class operator $C_{\a} \otimes C_{\beta}$
and density operator $ \rho_A \otimes \rho_B $, the linear positivity condition is
\beq
{\rm Re} \left( {\rm Tr} \left( C_{\a} \rho_A \right) {\rm Tr} \left( C_{\b} \rho_B \right) \right)\ge 0
\label{LP2}
\eeq
In any sensible assignment of probabilities, one would expect the two components to be statistically independent, in which case Eq.(\ref{LP2}) will be logically equivalent to two conditions of the form Eq.(\ref{LP1}). This requirement is known as the Diosi test and clearly fails for linear positivity. (We do not necessarily rule out the possibility that there are interesting and consistent extensions of quantum theory which do not reflect subsystem independence and this possibility is an interesting question for future research. However, it is clearly of interest to find situations which pass the test).

Our procedure of assigning probabilities by matching to marginals in fact passes the Di\'osi test.
This is easily seen. Suppose we have found a set of probabilities $p_A (\an)$ for system $A$, which match a given set of marginals, of the form $ p_A (\a_j, \a_k) $ for example, and probabilities $p_B (\bn)$ for system $B$ which match a given set of marginals $p_B (\b_{\ell},\b_m)$, say. Now we look at the composite system consisting of $A$ and $B$ and ask for the joint probabilities matching all the marginals of
the form $ p_A (\a_j, \a_k) p_B ( \b_{\ell}, \b_m) $. The solution is clearly the probability
\beq
p_{AB} (\an, \bn) = p_A (\an) p_B(\bn)
\eeq
The assignment of probabilities to the composite system is therefore logically equivalent to the assignment of probabilities to the two component systems.

It is interesting to link this to the Goldstein-Page formula. Suppose that linear positivity is satisfied for each subsystem. Expressions of the form $ {\rm Re} {\rm Tr} \left( C_{\a} \rho \right) $ will therefore arise as possible probabilities when we follow our procedure of matching to marginals. Now when we attempt to find probabilities for the composite system, we will find families of probabilities which will include the expression,
\beq
p(\an, \bn ) = {\rm Re} {\rm Tr} \left( C_{\a} \rho_A \right) {\rm Re} {\rm Tr} \left( C_{\b} \rho_B \right)
\eeq
This is actually the intuitively desired answer, although it does not follow from strict application of the linear positivity scheme, as we have indicated. The reason this seems to work out is that it has arisen from the application of a general {\it procedure} applied to probabilities, rather than a specific formula.

Our approach, however, does have the disadvantage that it generates a whole family of probabilities. This and related issues will be explored in future publications.

\section{Summary and Conclusions}

The purpose of this paper was to elucidate the underlying structures that might be relevant to the intuitive use and interpretation of quasi-probabilities, especially those of the type that arise in quantum mechanics for non-commuting variables.
Our main result is that quasi-probabilities may be divided into two qualitatively different types: viable quasi-probabillities, whose positive marginals are those of a positive probability, and non-viable ones, whose positive marginals cannot be matched to a positive probability. We gave some arguments to indicate that viable
quasi-probabilities are qualitatively similar to probabilities, so may enjoy an interpretation
similar to that of positive probabilities. For non-viable quasi-probabilities, we
suggested that this is not the case.


The question of matching a probability to a given set of marginals is a non-trivial
one in general. We showed that it may be solved in simple examples using the Bell
and CHSH inequalities, coupled with Fine's theorem. We also showed how much more
complicated examples could be boiled down to these simple cases by coarse graining,
leading to a method for proving non-viability for a wide variety of quasi-probabilities.
However, viability is very difficult to establish in general.

We considered the linear positivity condition of Goldstein and Page in the light of
our results. Our viability test is generally weaker than the linear positivity condition
and we demonstrated this in the particular example of probabilities for measurements
on the EPRB state. For certain detector orientations linear positivity can fail
when the CHSH inequalities are satisfied, so an underlying probability exists, but it is not the Goldstein-Page formula. This has the interesting consequence that an
underlying probabilty exists even though the weakest known condition for the assignment
of probabilities to non-commuting variables fails.

Our results have some implications for the histories approach to quantum
theory. They indicate that it would be of interest to undertake a search for the
most general possible method for the assignment of probabilities to histories, matching
some simple reasonable requirements. In particular we have argued that our method gives a way of finding families of probabilities for histories, with the important property that they pass the Di\'osi test. This will be explored further elsewhere.

\section{Acknowledgements}

We are grateful to Jim Hartle and Fay Dowker for useful conversations.
JJH was supported by a grant from EPSRC.
JMY was supported by a grant from the John Templeton Foundation.


\bibliography{apssamp}

\end{document}